\documentclass[10pt,twocolumn,aps,american,pre,showpacs,preprintnumbers,amsmath,amssymb]{revtex4-1}

\usepackage{babel} 
\usepackage[latin9]{inputenc}
\usepackage{lineno,hyperref}
\usepackage{amsbsy}
\usepackage{amstext}
\usepackage{amssymb}
\usepackage{graphicx}
\usepackage{esint}
\usepackage{amsmath}
\usepackage[all]{xy}
\usepackage{multirow,array}
\usepackage{xcolor}
\usepackage{float}
\usepackage{units}
\usepackage{multirow}
\usepackage{array}
\usepackage[titletoc]{appendix}
\usepackage{rotating}
\usepackage{dcolumn}
\usepackage{bm}
\usepackage{fancyhdr} 
\usepackage{braket}						
\usepackage{color} 							
\usepackage{lastpage} 						
\usepackage{hyperref} 						
\usepackage{url}
\usepackage{bibentry}
\usepackage{multirow}
\usepackage{silence}
\usepackage{appendix}
\usepackage{natbib}

\usepackage{tikz}
\usetikzlibrary{shapes}

\begin{document}

\newcommand{\entdash}{\raisebox{0pt}{\tikz[baseline=-0.6ex]{\draw[thick,dashed] (0,0)--(0.54,0);}}}
\newcommand{\pauliv}{\raisebox{0pt}{\tikz{\node[draw,scale=0.3,regular polygon, regular polygon sides=3,fill=none,red,rotate=180](){};}}}
\newcommand{\transv}{\raisebox{0pt}{\tikz{\node[draw,scale=0.4,regular polygon, regular polygon sides=4,fill=none,blue!50,rotate=180](){};}}}
\newcommand{\reflv}{\raisebox{0pt}{\tikz{\node[draw,scale=0.35,diamond,fill=none,green!40!black,rotate=180](){};}}}

\newcommand{\entropia}{\raisebox{0.5pt}{\tikz{\node[scale=0.35,circle,fill=black](){};}}}

\newcommand{\pauliI}{\raisebox{0pt}{\tikz{\node[scale=0.3,regular polygon, regular polygon sides=3,fill=red,rotate=180](){};}}}
\newcommand{\transI}{\raisebox{0pt}{\tikz{\node[scale=0.4,regular polygon, regular polygon sides=4,fill=blue!50,rotate=0](){};}}}
\newcommand{\reflI}{\raisebox{0pt}{\tikz{\node[scale=0.35,diamond,fill=green,rotate=0](){};}}}

\newcommand{\pauliNI}{\raisebox{0pt}{\tikz{\node[draw,scale=0.3,regular polygon, regular polygon sides=3,fill=none,red!40!black,rotate=180](){};}}}
\newcommand{\transNI}{\raisebox{0pt}{\tikz{\node[draw,scale=0.4,regular polygon, regular polygon sides=4,fill=none,blue!40!black,rotate=0](){};}}}
\newcommand{\reflNI}{\raisebox{0pt}{\tikz{\node[draw,scale=0.35,diamond,fill=none,green!40!black,rotate=0](){};}}}

\renewcommand{\thetable}{\arabic{table}}

\title{Relevant OTOC operators: footprints of the classical dynamics}

\author{Pablo D. Bergamasco}
\affiliation{Departamento de F\'isica, CNEA, Libertador 8250, (C1429BNP) Buenos Aires, Argentina}
\author{Gabriel G. Carlo}
\author{Alejandro M. F. Rivas}
\affiliation{Departamento de F\'isica, CNEA, CONICET, Libertador 8250, (C1429BNP) Buenos Aires, Argentina}

\date{\today}

\begin{abstract}
The out-of-time order correlator (OTOC) has recently become relevant in different areas 
where it has been linked to scrambling of quantum information and entanglement. It has also 
been proposed as a good indicator of quantum complexity. In this sense, the OTOC-RE theorem 
relates the OTOCs summed over a complete base of operators to the second Renyi entropy. 
Here we have studied the OTOC-RE correspondence on physically meaningful bases like the 
ones constructed with the Pauli, reflection, and translation operators.
The evolution is given by a paradigmatic bi-partite system consisting of two perturbed and 
coupled Arnold cat maps with different dynamics.
We show that the sum over a small set of relevant operators, is enough in order to obtain a very 
good approximation for the entropy and hence to reveal the character of the dynamics, 
up to a time $t_0$. In turn, this provides with an alternative natural indicator of complexity, 
i.e. the scaling of the number of relevant operators with time. When represented in phase space, 
each one of these sets reveals the classical dynamical footprints with different depth according 
to the chosen base.

\end{abstract}


\maketitle

\section{Introduction}
\label{sec:intro}

There is a great interest in the OTOC nowadays, coming from different areas like high energy and gravity, condensed matter, many-body systems, quantum information, and quantum chaos. This measure has been introduced in the superconductivity context \cite{larkin1969quasiclassical} where the exponential growth as a function of time has been associated with chaotic behavior. The OTOC is usually defined as a 4-point out-of-time order correlator,
\begin{equation}
  C(t) = \langle \hat{M}(t)\hat{V}(0)\hat{M}(t)\hat{V}(0)\rangle
  \label{eq:OTOC}
\end{equation}
where $\langle \cdot \rangle = Tr[\cdot]/N$ is the thermal average and $M(t)$ is an operator evolved in the Heisenberg picture. The establishing of an upper limit to the growth rate of the OTOC in black hole models \cite{maldacena2016bound} has led to an interest surge on this versatile 
measure. Examples of this can be found in many-body physics 
\cite{shenker2014black, aleiner2016microscopic, huang2017out, borgonovi2018emergence, 
slagle2017out, chen2017out, Richter, garcia20}, quantum chaos 
\cite{lakshminarayan2018out, garcia2018chaos, jalabert2018semiclassical}, high energy physics \cite{Watanabe}, and the link between topological gravity and quantum chaos \cite{liu20}. Recently, the OTOC behavior has been studied for bi-partite systems. In \cite{prakash2019scrambling} it was found that for the chaotic case the scrambling process has two phases, one in which the exponential growth 
is within the subsystem and a second one which depends only on the interaction. In \cite{PhysRevResearch.1.033044} the OTOC has proven to be a very good indicator of quantum complexity \cite{bergamasco2017,Benenti-Carlo-Prosen} when considering all possible dynamical scenarios.

The OTOC is conceptually related to scrambling of quantum information \cite{pappalardi2018scrambling, campisi2017thermodynamics, swingle2018unscrambling} and entanglement \cite{PhysRevResearch.1.033044}. It is in this respect that the OTOC-RE theorem \cite{hosur2016chaos,fan2017out} establishes the equivalence of the linear entropy $S_{L}$ with the 4-point OTOC averaged over a complete operator basis of some arbitrary partition of the system.
Following the scheme presented in \cite{fan2017out}, we can summarize the theorem as
\begin{equation}
  S_{L}=1-e^{(-S^{(2)}_{A})} = 1 - \sum_{\hat{M}\in B}{Tr[\hat{M}(t)\hat{\rho}(0)\hat{M}^{\dagger}(t)\hat{\rho}(0)]}
\label{eq:OTOC-RE}
\end{equation}
where $A$ and $B$ are two partitions of our system, $\hat{\rho}(0)$ is the initial (non evolving) density operator of the whole system, $S^{(2)}_{A} = -log Tr[\hat{\rho}^{2}_{A}]$ is the second Renyi entropy and $S_{L}$ is the linear entropy. The $\hat{M}$ operators act on the subsystem $B$ and define a complete basis normalized according to $\sum_{\hat{M}\in B}{M_{ij}(M_{lm})^{\dagger}}=\delta_{im}\delta_{lj}$. In Eq.\ref{eq:OTOC-RE} we have taken the second evolved operator as $\hat{M}^{\dagger}(t)$, being the transpose and conjugate of the first one in such a way to extend the validity of the theorem to unitary operators. 
This result prescribes an average behavior for different OTOCs in a given basis, but it is 
important to ask ourselves how meaningful this is. As a matter of fact, is each one of the terms appearing in \ref{eq:OTOC-RE} equally relevant, making the 
same contribution to the linear entropy? In this work we determine that not all of the OTOCs are 
good indicators of quantum complexity, but we are able to classify them in terms of the information they provide on the dynamical features.

Our system consists of two perturbed and coupled Arnold cat maps with different dynamics. The three possible cases were considered, i.e. both maps being hyperbolic (chaotic) (HH), both elliptic (regular) (EE) and a mixed scenario where one map is hyperbolic and the other is elliptic  (HE,EH) \cite{PhysRevResearch.1.033044}. Also, we have considered three different bases constructed with Pauli or $SU(N)$, translation and reflection operators on the torus \cite{rivas1999weyl}. In all cases we have taken the non evolving density operator as localized pure states.
Our results show that performing the summation in Eq.\ref{eq:OTOC-RE} with a set of only $35\%$ or less of the operators, in any of the chosen basis, $80\%$ of $S_{L}$ is recovered. On the other hand, this set of relevant operators is given by those that best capture the dynamics of the system, being suitable for complexity measures. For reflection and translation bases, they show clear 
footprints of the underlying classical dynamics in phase space.

This paper is organized as follows: in Section \ref{sec:model} we present our system with a brief description of the properties of the Hilbert space on the torus. We also describe the operator bases that we use for the OTOC-RE theorem analysis. In Section \ref{sec:result} we explain our results in detail and in Section \ref{sec:conclusion} we state our conclusions.

\section{System and Bases}
\label{sec:model}

The periodicity of the torus implies Bloch boundary conditions for wave functions:
\begin{eqnarray}
  \Psi(q + 1) = e^{2\pi i \chi_{p}} \Psi(q)\nonumber\\
  \tilde{\Psi}(p + 1) = e^{2\pi i \chi_{q}} \tilde{\Psi}(p)\nonumber
\label{eq:boundary}
\end{eqnarray}
where
\[
  \tilde{\Psi}(p) = \frac{1}{\sqrt{2\pi\hbar}} \int{e^{-ipq/\hbar}} \Psi(q) dq
\]
with $2\pi i \chi_{p}$ and $2\pi i \chi_{q}$ arbitrary Floquet angles that determine the so called prequantization. The values of $\chi_{p},\chi_{q}$ can be chosen in the range $[0,1]$, 
we take $\chi_{p} = \chi_{q} = 0$. The previous boundary conditions can be satisfied if 
there is an integer $N$, so that\cite{bouzouina1996equipartition}
\begin{equation}
  \hbar = \frac{1}{2\pi N}.
\label{eq:hbarN}
\end{equation}
This implies a Hilbert space $\mathcal{H}_{N}$ of finite dimension $N$. 
We take $\ket{q_n}$ and $\ket{p_m}$ with $n,m=0,1,\dots, N-1$ as bases of $\mathcal{H}_{N}$. 
The states $\langle q|q_j\rangle$ are periodic Dirac delta distributions 
at positions $q=n/N{\rm mod}(1)$, with $n$ an integer in $[0,N-1]$.
These bases have the following normalization conditions,
\[
  \langle{q_{m}}|{q_{n}}\rangle = \langle{p_{m}}|{p_{n}}\rangle = \delta^{(N)}_{m,n}
\]
with $\delta^{(N)}_{i,j}$ the N-periodic Kronecker delta defined as
\[
  \delta^{(N)}_{i,j} =  \sum_{k=-\infty}^{\infty} \delta_{i,j+kN}
\]
The bases are exchanged with the transformation kernel,
\[
  \langle{p_{m}}|{q_{n}}\rangle = \frac{1}{\sqrt{N}} e^{\frac{2\pi imn}{N}}.
\]
Position and momenta are then points in a discrete lattice on the torus with separation 
$1/N$, i.e. the quantum phase space \cite{galetti1988extended}.

The quantization of the cat map \cite{Hannay1980}  which is one of the most simple paradigmatic  models of chaotic dynamics,  has helped to elucidate many questions in the quantum chaos area \cite{Hannay1980,Ozorio1994,Haake,Espositi 2005}.
Here we consider the behavior of two coupled perturbed cat maps, a two degrees of freedom example, 
which can have different types of dynamics.
For each degree of freedom, the map is defined on the 2-Torus as \cite{Hannay1980}
\begin{equation}
  \begin{pmatrix}
    q_{t+1}\\
    p_{t+1}
  \end{pmatrix} = M
  \begin{pmatrix}
    q_{t}\\
    p_{t} + \epsilon(q_{t})
  \end{pmatrix}
\label{eq:mapa1D}
\end{equation}
with $q$ and $p$ taken modulo 1, and the perturbation
\[
  \epsilon(q_{t}) = -\frac{K}{2\pi} \sin{(2\pi q_{t})}.
\]
The matrix $M$ defines the dynamics. For the chaotic case we have chosen the hyperbolic map
\begin{equation}
  M_{h} =
  \begin{pmatrix}
  2&1 \\
  3&2
  \end{pmatrix},
  \label{eq:Mhiperbolica}
\end{equation}
while for the regular behavior we have taken the elliptic map
\begin{equation}
  M_{e} =
  \begin{pmatrix}
  0 & 1 \\
  -1&0
  \end{pmatrix}.
  \label{eq:Meliptica}
\end{equation}
The propagator in position representation is given by the $N\times N$ unitary matrix
\begin{equation}
  U_{jk}=A{\exp}\left[\frac{i\pi}{N\mathcal{M}_{12}}(\mathcal{M}_{11}j^{2}-2jk +\mathcal{M}_{22}k^{2})+F\right]
\label{eq:U1D}
\end{equation}
where
\[
A=[1/\left(iN\mathcal{M}_{12}\right)]^{1/2}
\]
and,
\[
F=[iKN/(2\pi)]\cos(2\pi j/N).
\]
We can extend it to two degrees of freedom defined in a four-dimensional phase space 
of coordinates $\left(q^{1},q^{2},p^{1},p^{2}\right)$ \cite{Benenti-Carlo-Prosen} as
\[
\left(\begin{array}{c}
q_{t+1}^{1}\\
p_{t+1}^{1}
\end{array}\right)=\mathcal{M}_{1}\left(\begin{array}{c}
q_{t}^{1}\\
p_{t}^{1}+\epsilon\left(q_{t}^{1}\right)+\kappa\left(q_{t}^{1},q_{t}^{2}\right)
\end{array}\right)
\]
and
\[
\left(\begin{array}{c}
q_{t+1}^{2}\\
p_{t+1}^{2}
\end{array}\right)=\mathcal{M}_{2}\left(\begin{array}{c}
q_{t}^{2}\\
p_{t}^{2}+\epsilon\left(q_{t}^{2}\right)+\kappa\left(q_{t}^{1},q_{t}^{2}\right)
\end{array}\right).
\]
where the coupling between both maps is given by $\kappa(q_{t}^{1},q_{t}^{2})$. 
Hence, the quantum evolution for this case is given by the tensor product of 
the one degree of freedom maps 
\[
U_{j_{1}j_{2},k_{1}k_{2}}^{2D}=U_{j_{1}k_{1}}U_{j_{2}k_{2}}C_{j_{1}j_{2}},
\]
with coupling matrix,
\[
C_{j_{1}j_{2}}=\exp\left\{ \left(\frac{iNK_{c}}{2\pi}\right)
\cos\left[\frac{2\pi}{N}\left(j_{1}+j_{2}\right)\right]\right\},
\]
where $j_{1},j_{2},k_{1},k_{2}\in\{0,\ldots,N-1\}$. We fix $K=0.25$ and $K_{c}=0.5$ 
(Anosov condition \cite{Ozorio1994}), and $N=2^6$ throughout this work.

For the complete set of operators spanning one of the subsystems in Eq. \ref{eq:OTOC-RE}  we have chosen three different sets. They are the so called computational or Pauli base, the translation base, and the reflection base \cite{rivas1999weyl}. The first one is relevant for multi-qubit systems 
(canonical in quantum computation and information), while the translation and reflection ones define the chord and Wigner (or center) functions \cite{ozrep} respectively, allowing for a more direct comparison with classical counterparts.

\textit{Pauli base.} For qubit systems, the typical base chosen is $\left\{ \sigma_{0},\sigma_{1},\sigma_{2},\sigma_{3}\right\}$ where $\sigma_{0} = \openone$ and the rest of the $\sigma_{i}$'s 
are $2 \times 2$ Pauli matrices. For dimensions $N=2^{k}$, we can extend this basis by taking the complete system as a direct product of $k$ single qubits,
\begin{equation}
  \left\{\bigotimes_{t = 1}^{k} \sigma_{j_{t}}\right\}
  \label{eq:SUbase}
\end{equation}

\textit{Translation and reflection bases.} In \cite{rivas1999weyl} translation operators 
$\hat{T}_\xi $ on the torus are described by their chord $\xi=(\xi_p,\xi_q) = (r/N,s/N)$ 
with $r$ and $s$ integer indices. A complete basis of $N^2$ independent operators is obtained for chords performing up to one loop on the torus, that is for $r$ and $s$ belonging to the interval $[0,N-1]$. The matrix elements of the translation operators $\hat{T}_\xi $ in the position representation are given by,
\begin{equation}
  \bra{q_{i}}\hat{T}_{\xi {\textit(r,s)}} \ket{q_{j}}= e^{i\frac{2\pi}{N}r(\frac{i+j}{2}+\chi_{q})} \delta^{(N)}_{j,i+s} e^{-i\frac{2\pi}{N}(\frac{r}{2}+\chi_{p})(j-i-s)}.
  \label{eq:TRAbase}
\end{equation}
For the case of reflection operators $\hat{R}_x $, they are described by their center point $x=(x_p,x_q) = (a/N,b/N)$ with half-integer indices $a$ and $b$ with values in $[0,\frac{N-1}{2}]$ in order to complete a basis of $N^2$ independent operators. That is, a quarter of the torus contains the complete information for the reflection basis. The matrix elements in the position 
representation are
\begin{equation}
  \bra{q_{i}} \hat{R}_{x {\textit(a,b)}} \ket{q_{j}} = e^{i\frac{2\pi}{N}(j-i)(a+\chi_{q})} \delta^{(N)}_{j,2b-i} e^{i\frac{2\pi}{N}a(2b-i-j)} .
\label{eq:REFbase}
\end{equation}
We recall that in both cases we have chosen the Floquet angles $(\chi_{q}, \chi_{p})$ as zero.

\section{Results}
\label{sec:result}

\begin{figure}[!htp]
\centering
\setlength{\tabcolsep}{-6.1pt} 
\setlength{\extrarowheight}{-7pt} 
  \begin{tabular}{cc}
  \includegraphics[scale=0.55]{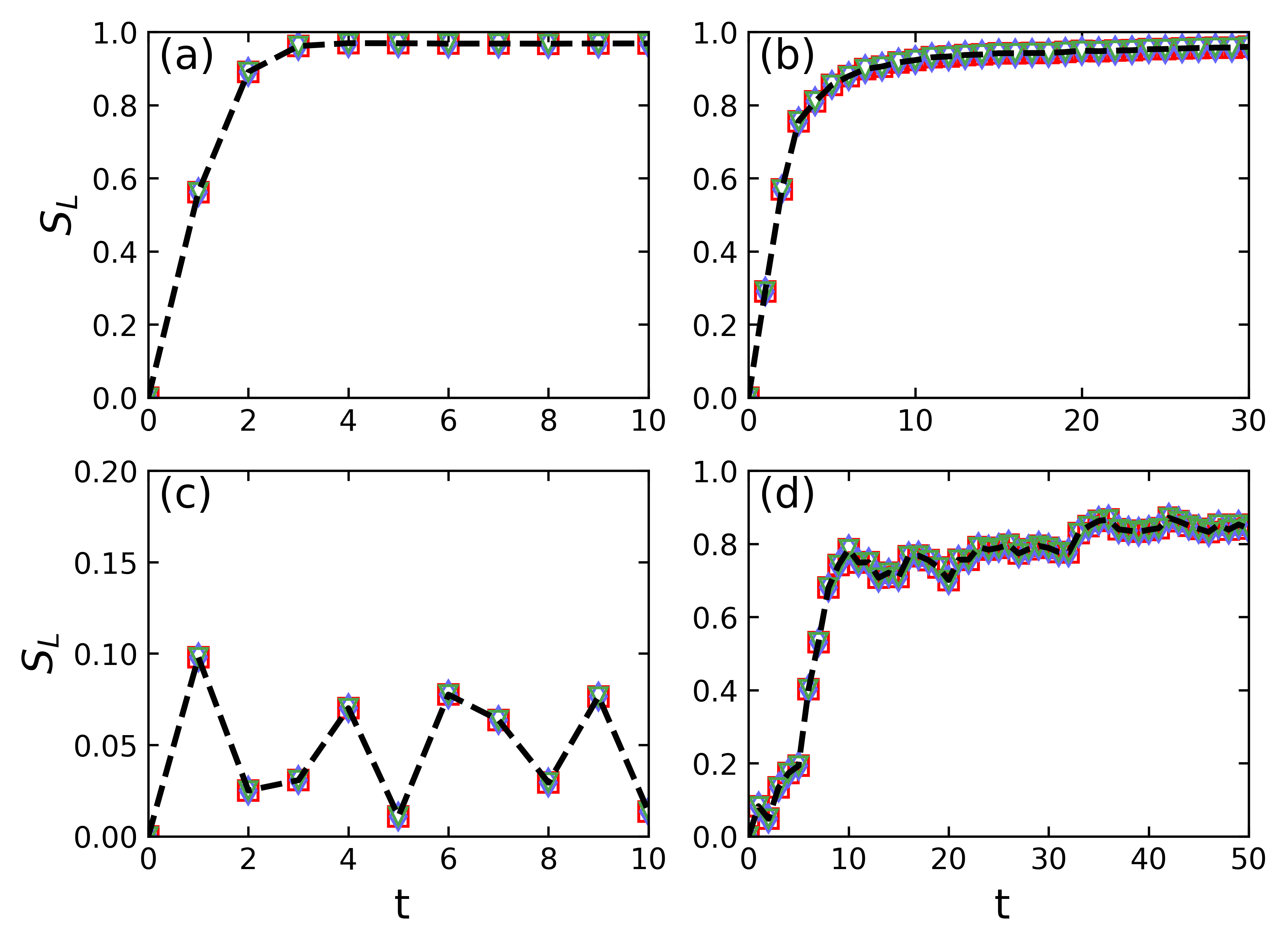}
\end{tabular}
  \caption{(Color online) In all panels we display $S_{L}$ ((black) dashed line) (L.H.S of Eq.\ref{eq:OTOC-RE}) and the complete sum of OTOCs (R.H.S of Eq.\ref{eq:OTOC-RE}) for the three different bases described in the main text: Pauli or $SU(N)$ ((red) squares),  translations ((blue) diamonds) and reflections ((green) down triangles). a) HH and b) HE cases. c) and d) EE cases with the coherent states at $(q,p)=(0.5,0.5)$ and $(q,p)=(\pi/4,\pi/4)$, respectively.}
  \label{fig1}
\end{figure}

For completeness, we first check the validity of the OTOC-RE theorem (Eq.\ref{eq:OTOC-RE}) for all the dynamical 
scenarios and all the operator bases described in Sec. \ref{sec:model}. Fig. \ref{fig1} a) corresponds to $S_{L}$ 
and all OTOCs sums as a function of the time $t$ (map steps), for both dynamics being hyperbolic (HH), 
while in Fig. \ref{fig1} b) and c) we show the HE and EE cases, respectively. We 
consider a coherent state located at the fixed point $(q,p)=(0.5,0.5)$ on each tori. 
Fig. \ref{fig1} d) displays the EE case where the coherent state is located at 
$(q,p)=(\pi/4,\pi/4)$ (not a fixed point). The theorem clearly holds regardless of the dynamics 
or the chosen base.

We have classified each OTOC in Eq. \ref{eq:OTOC-RE} according to its contribution to the sum. In fact, their relevance is 
essentially given by the corresponding area under the curve up to a time $t_{0}$.
We proceed in the following way: for each operator $M$ we have calculated the area $A_M(t_0)$ as 
\begin{equation}
  A_M(t_0) = \int_{0}^{t_0}{C_M(t)dt}.
\end{equation}
where $C_M(t)$ is the OTOC 
\begin{equation}
C_M(t) = Tr[\hat{M}(t)\hat{\rho}(0)\hat{M}^{\dagger}(t)\hat{\rho}(0)].
\label{eq:RElev}
\end{equation}
Then, we have ordered the operators using $A_M(t_0)$ which reflects their contribution to the total area under $1-S_{L}(t)$, 
given by $A_S(t_0) = \int_{0}^{t_0}{1-S_{L}(t)dt}$. Finally we determine a cutoff criterion which consists of reaching 
the value $0.8 A_S(t_0)$ by simply adding the areas contributed by each operator's OTOC like $\sum_{R}A_M(t_0)$, 
where $R$ means that the sum only runs from the most up to the least relevant one. This provides us with the number 
of OTOCs necessary to reach what we will refer to as the effective $S_{L}$ behavior.

\begin{figure}[!htp]
\centering
  \includegraphics[scale=0.5]{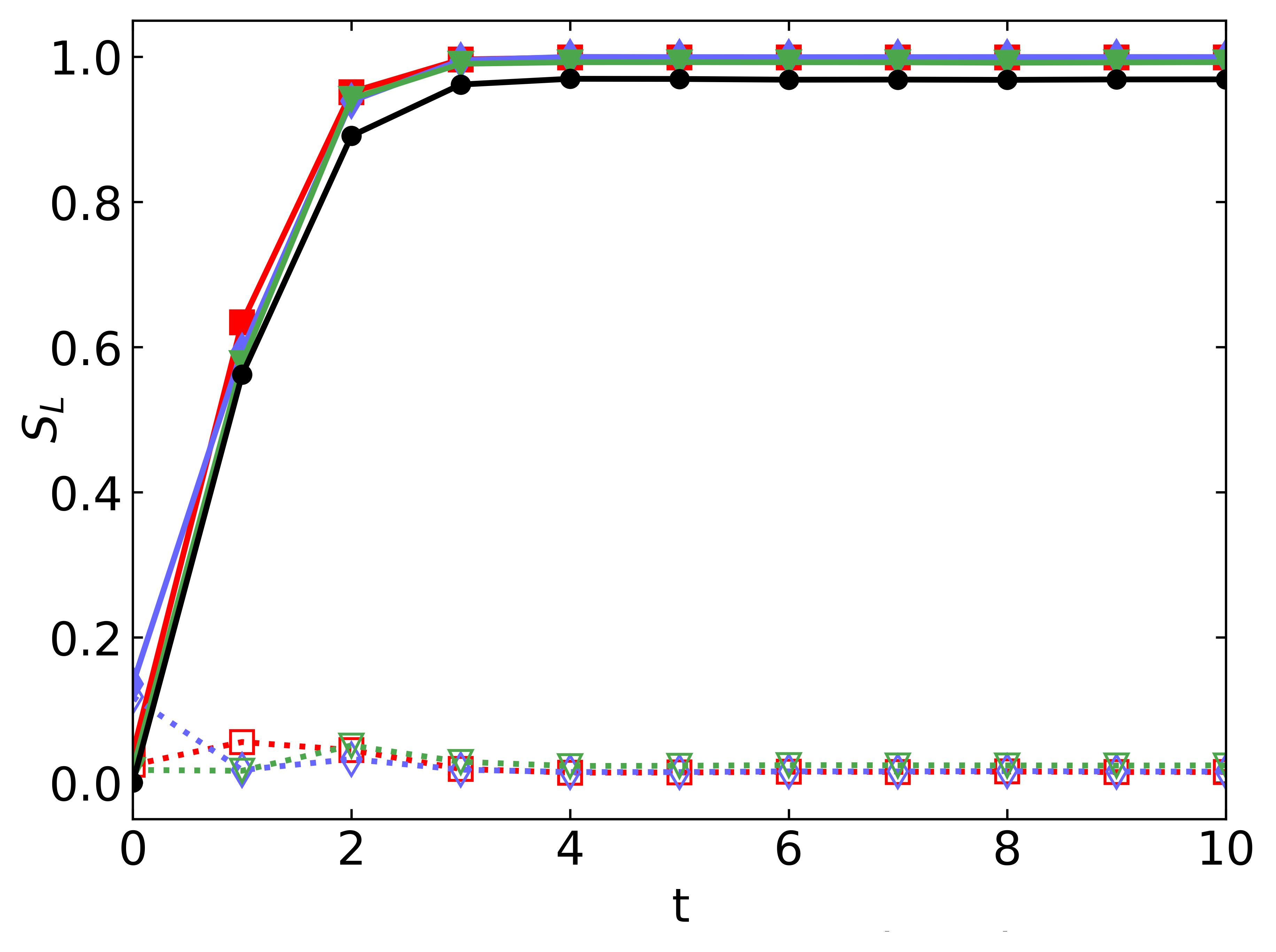}
  \caption{(Color online) $S_{L}$ ((black) solid line with filled circles) given by the l.h.s of Eq.\ref{eq:OTOC-RE} for the HH case with coherent states taken 
  at $(q,p)=(0.5,0.5)$. Pauli ((red) solid line with filled squares) operators sum considering the $263$ most relevant terms in the r.h.s of Eq.\ref{eq:OTOC-RE}. Translation ((blue) solid line with filled diamonds) and reflection ((green) solid line with filled down triangles) bases with $166$ and $697$ 
  terms respectively. Empty symbols with dotted lines show the contribution of the remaining terms for 
  the Pauli ((red) dashed line with empty squares), translation ((blue) dashed line with empty diamonds) and reflection ((green) dashed line with empty down triangles) bases. $t_0=10$.}
  \label{fig2}
\end{figure}

We first consider the HH case with coherent states at $(q,p)=(0.5,0.5)$. Due to the chaotic nature of the 
dynamics the OTOCs and $S_{L}$ both grow exponentially \cite{PhysRevResearch.1.033044} at an early stage, hence we only 
look up to $t_0=10$. In Fig. \ref{fig2} we show $S_{L}(t_0)$ (black lines) and the partial sum obtained with the 
most relevant OTOCs (filled symbols) for each operator base. For the Pauli base, only $263$ from a total of $4096$ terms 
were needed in order to reach the effective $S_{L}$ behavior. Meanwhile, for the translation and reflection bases $166$ and $697$ terms were 
needed, respectively. The effective entropy behavior is recovered with less than $20\%$ of the operators. 
In addition, in Fig \ref{fig2} we also show the contribution of the remaining OTOCs (empty symbols) which is markedly 
lower than that of the most relevant ones. We notice that in all Figures we display $1-\sum_{R}{C_M(t)}$ which is directly 
compared to $S_{L}$, then the values corresponding to the empty symbols 
are to be subtracted from the filled ones to recover the entropy.

\begin{figure}[!htp]
\centering
  \includegraphics[scale=0.5]{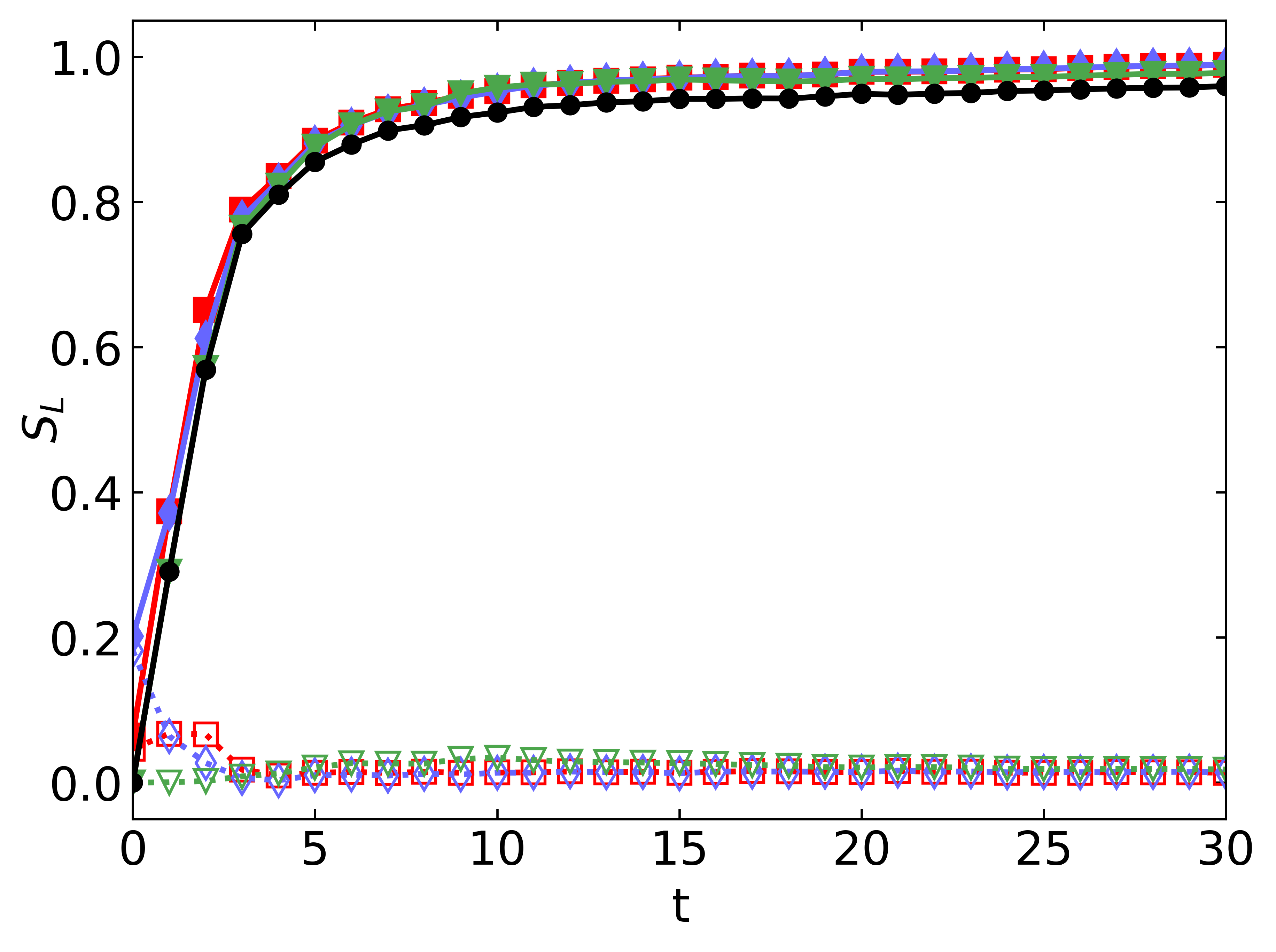}
    \caption{(Color online) $S_{L}$ ((black) solid line with filled circles) given by the l.h.s of Eq.\ref{eq:OTOC-RE} for the HE case with coherent states taken 
  at $(q,p)=(0.5,0.5)$. Pauli ((red) solid line with filled squares) operators sum considering the $199$ most relevant terms in the r.h.s of Eq.\ref{eq:OTOC-RE}. Translation ((blue) solid line with filled diamonds) and reflection ((green) solid line with filled down triangles) bases with $110$ and $1285$ 
  terms respectively. Empty symbols with dotted lines show the contribution of the remaining terms for 
  the Pauli ((red) dashed line with empty squares), translation ((blue) dashed line with empty diamonds) and reflection ((green) dashed line with empty down triangles) bases. 
  $t_0=40$.}
  \label{fig3}
\end{figure}

Next we look into the HE map where the operator base is taken for the regular subsystem and coherent states are placed at 
$(q,p)=(0.5,0.5)$. In Fig\ref{fig3} we see that $S_{L}$  grows slower than in the previous case (until saturation) 
due to the mixed character of the dynamics, leading us to a longer integration time ($t_0=40$). 
To recover the effective $S_{L}$ behavior this time we needed $199$ Pauli, $110$ translation and $1285$ 
reflection operators, i.e. less than $30\%$ of the total operators in the worst case. 

\begin{figure}[!htp]
\centering
  \includegraphics[scale=0.5]{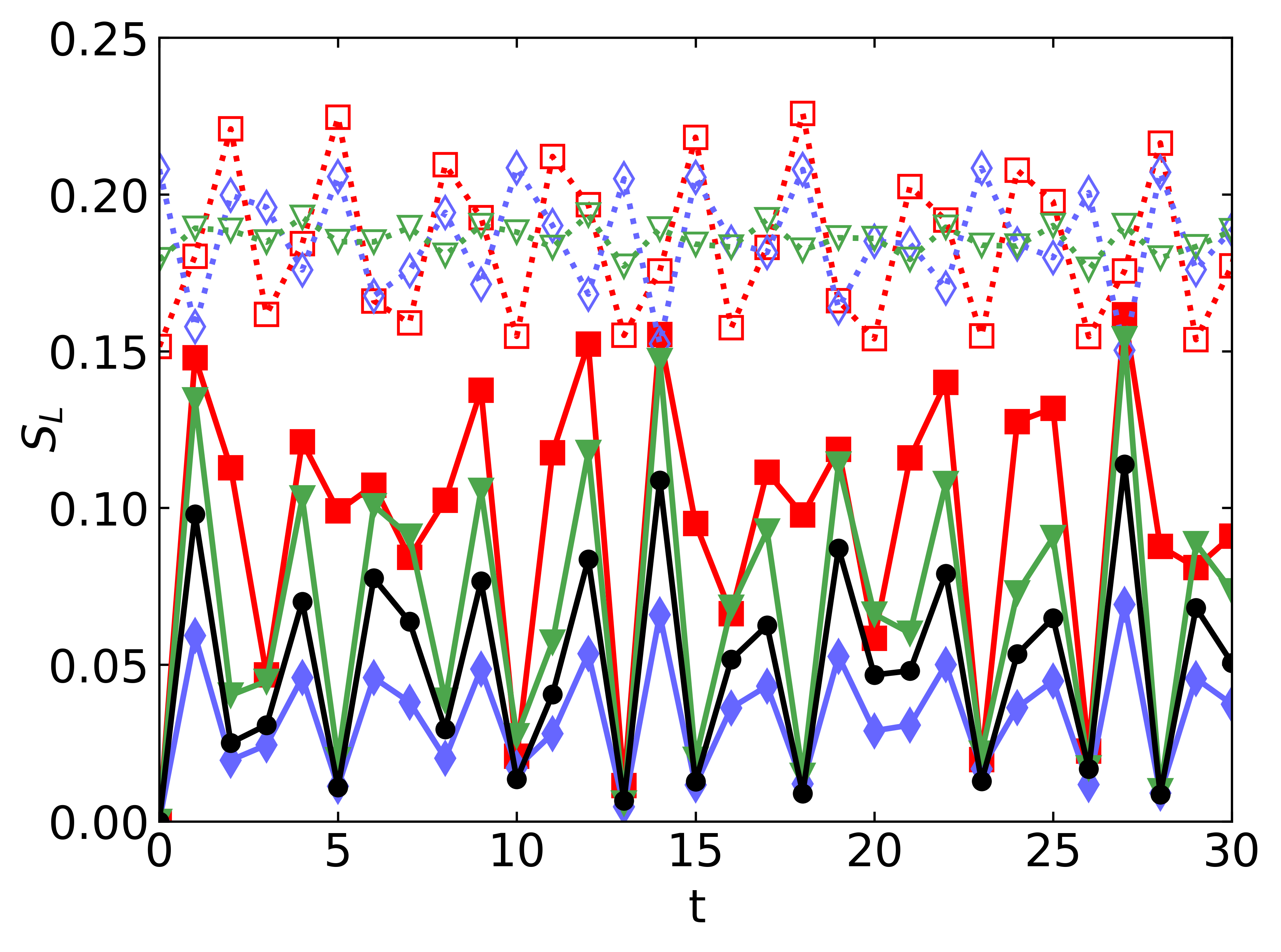}
    \caption{(Color online) $S_{L}$ ((black) solid line with filled circles) given by the l.h.s of Eq.\ref{eq:OTOC-RE} for the EE case with coherent states taken 
  at $(q,p)=(0.5,0.5)$. Pauli ((red) solid line with filled squares) operators sum considering the $81$ most relevant terms in the r.h.s of Eq.\ref{eq:OTOC-RE}. Translation ((blue) solid line with filled diamonds) and reflection ((green) solid line with filled down triangles) bases with $101$ and $117$ 
  terms respectively. Empty symbols with dotted lines show the contribution of the remaining terms for 
  the Pauli ((red) dashed line with empty squares), translation ((blue) dashed line with empty diamonds) and reflection ((green) dashed line with empty down triangles) bases. 
  $t_0=10$.}
  \label{fig4}
\end{figure}

\begin{figure}[!htp]
\centering
  \includegraphics[scale=0.5]{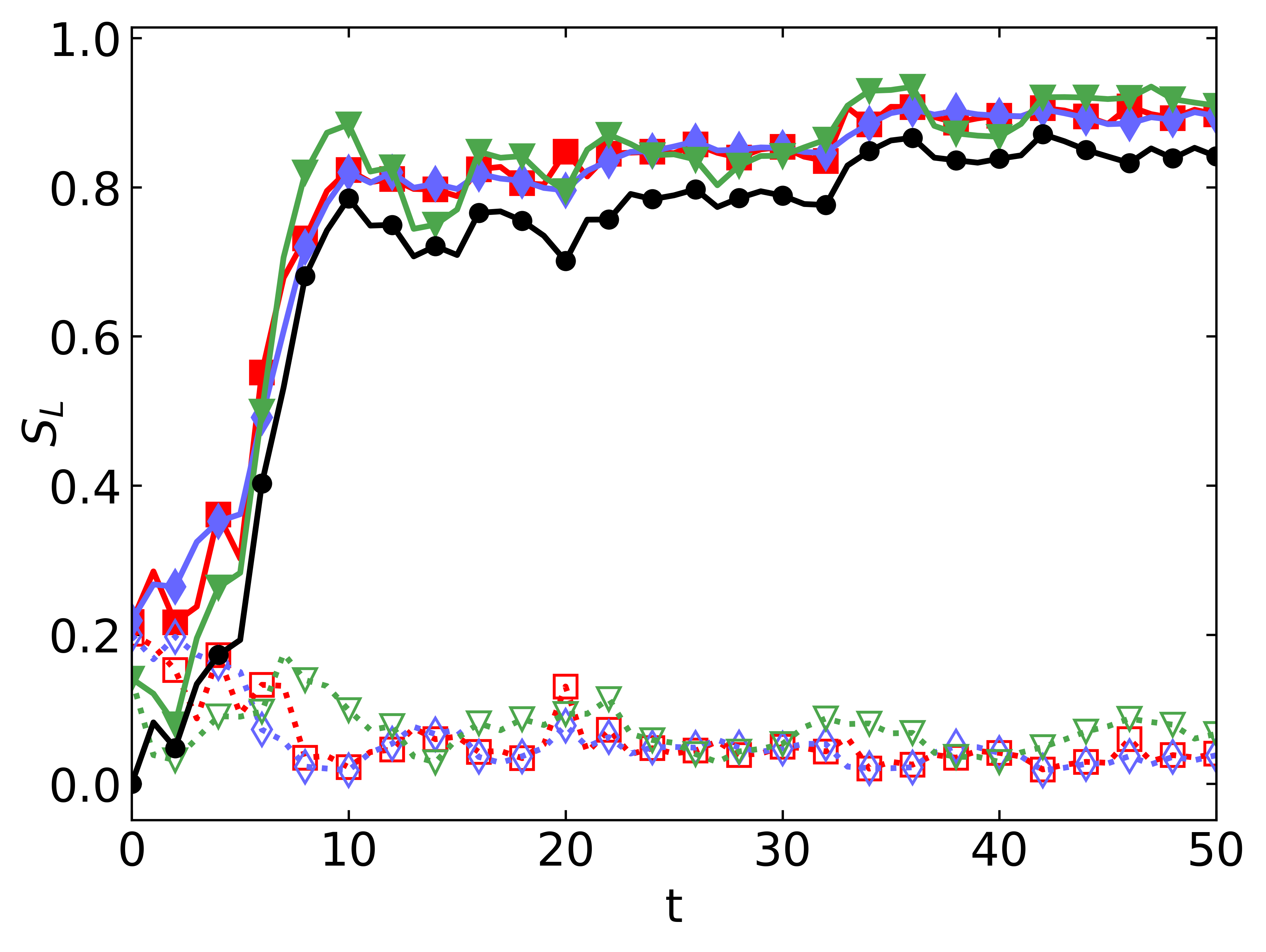}
  \caption{(Color online) $S_{L}$ ((black) solid line with filled circles) given by the l.h.s of Eq.\ref{eq:OTOC-RE} for the EE case with coherent states taken 
  at $(q,p)=(\pi/4,\pi/4)$. Pauli ((red) solid line with filled squares) operators sum considering the $413$ most relevant terms in the r.h.s of Eq.\ref{eq:OTOC-RE}. Translation ((blue) solid line with filled diamonds) and reflection ((green) solid line with filled down triangles) bases with $103$ and $1450$ 
  terms respectively. Empty symbols with dotted lines show the contribution of the remaining terms for 
  the Pauli ((red) dashed line with empty squares), translation ((blue) dashed line with empty diamonds) and reflection ((green) dashed line with empty down triangles) bases. 
  $t_0=30$.}
  \label{fig5}
\end{figure}

Finally, we take the EE map with coherent states at $(q,p)=(0.5,0.5)$ and then at $(q,p)=(\pi/4,\pi/4)$. The first case is 
shown in Fig. \ref{fig4} ($t_0=10$), where the effective $S_{L}$ behavior is recovered by just $81$ operators in the Pauli, $101$ in the translation, and $117$ in the reflection bases. In this dynamical scenario the coherent state does not explore the entire phase space but just rotates around the fixed point, giving a hint to explain this clear reduction in the number of relevant operators. In this case we have re-scaled the partial sum of the most relevant OTOCs for a better comparison with $S_{L}$ (the sum of the remaining ones is left unchanged). 
In Fig. \ref{fig5} we display the results when placing the coherent states at $(q,p)=(\pi/4,\pi/4)$. Since this is not at a fixed point the entropy grows up to saturation at a slower rate than in the HH case, leading us to consider $t_0=30$. We recover the effective $S_{L}$ behavior with $413$ operators in the Pauli base, $103$ in the translation and $1450$ in the reflection one, i.e. about $35\%$ of the operators in the worst case.

We mention that not only the sum but each one of the quantities $1-C_{M_{R}}(t)$ (where $\hat{M}_R$ stands for the 
relevant operators) approximate the linear entropy very well (up to normalization), i.e. we claim that 
$S_{L}=1-e^{(-S^{(2)}_{A})} \approx 1 - C_{M_{R}}(t)$. The remaining operators have a different behavior. 

\begin{figure}[htp]
\begin{center}
  \hspace*{-1cm}
  \includegraphics[scale=0.5]{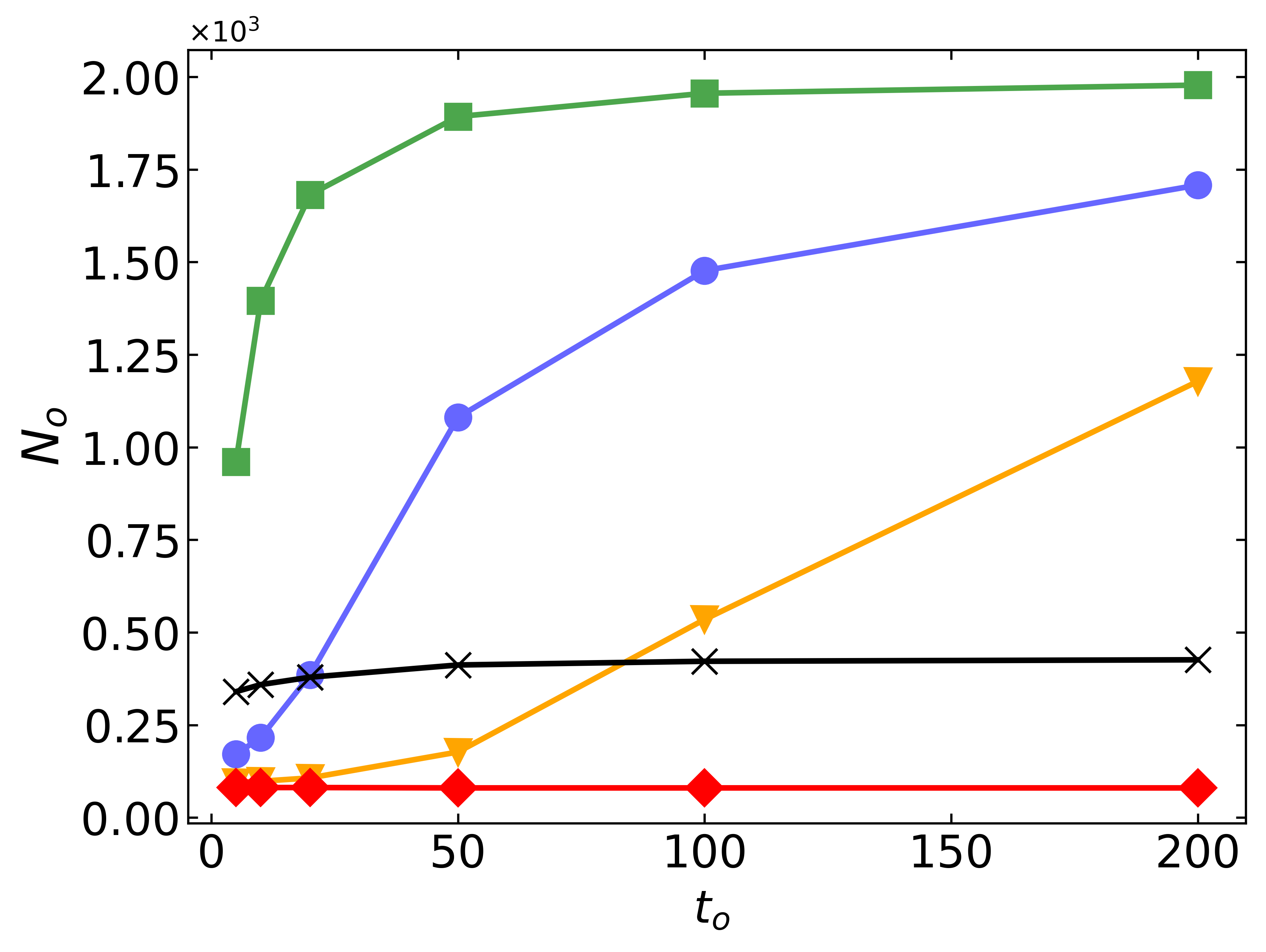}
  \caption{(Color online) Number of relevant Pauli operators, for  different integration times $t_{0}$ and dynamics. (Blue) line with circles stands for the HH case, (green) line with squares for EH, (orange) line with down triangles for HE, and 
  finally (red) line with diamonds and (black) line with crosses for the EE cases with coherent states at $(0.5, 0.5)$ and $(\pi/4, \pi/4)$, respectively. $N=2^{6}$.}
  \label{fig6}
\end{center}
\end{figure}

\begin{figure}[htp]
\centering
\hspace*{-1cm}
  \includegraphics[scale=0.5]{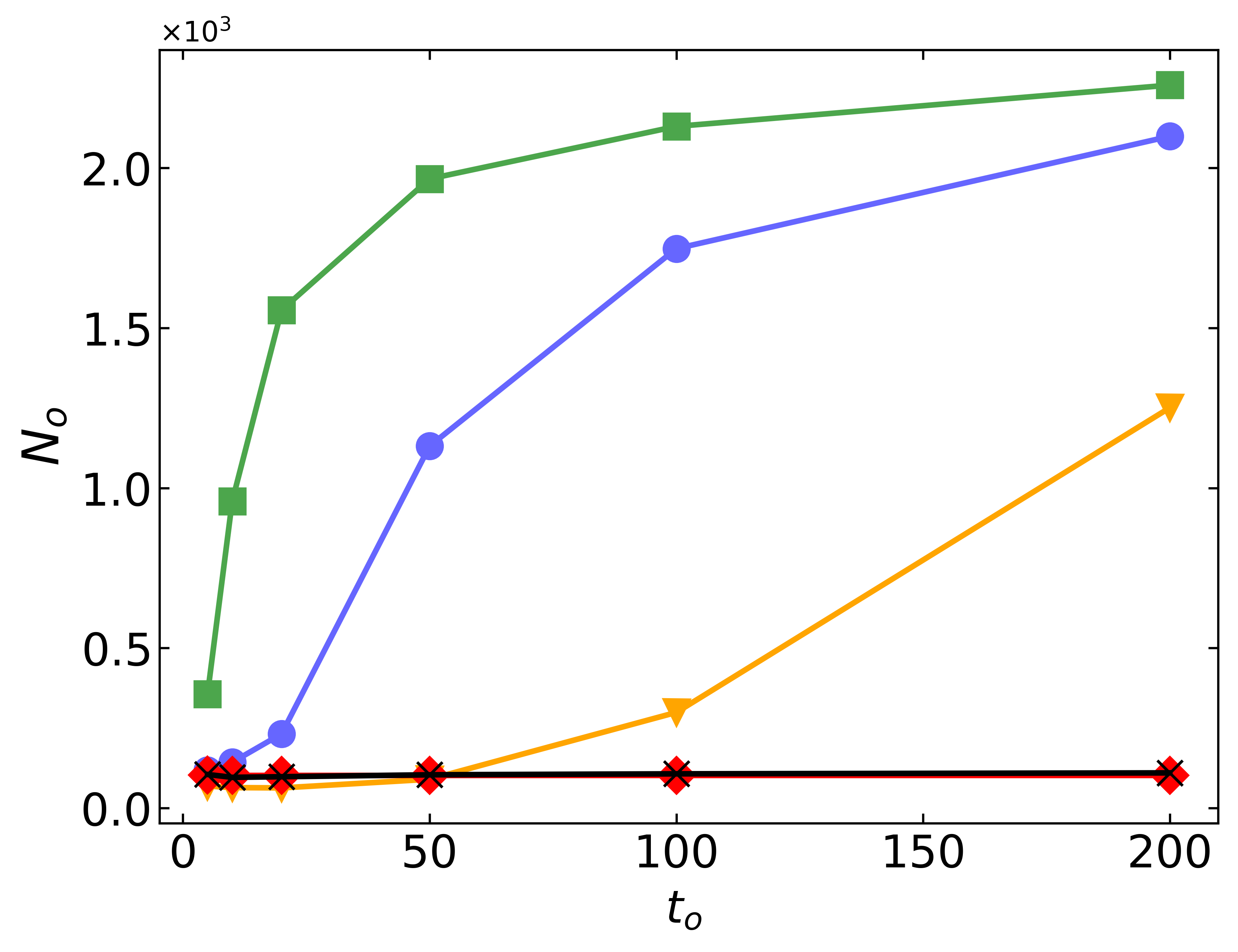}
  \caption{Number of relevant translation operators, for  different integration times $t_{0}$ and dynamics. Same color code as in Fig. \ref{fig6}. $N=2^{6}$.}
  \label{fig7}
\end{figure}

\begin{figure}[htp]
\centering
\hspace*{-1cm}
  \includegraphics[scale=0.5]{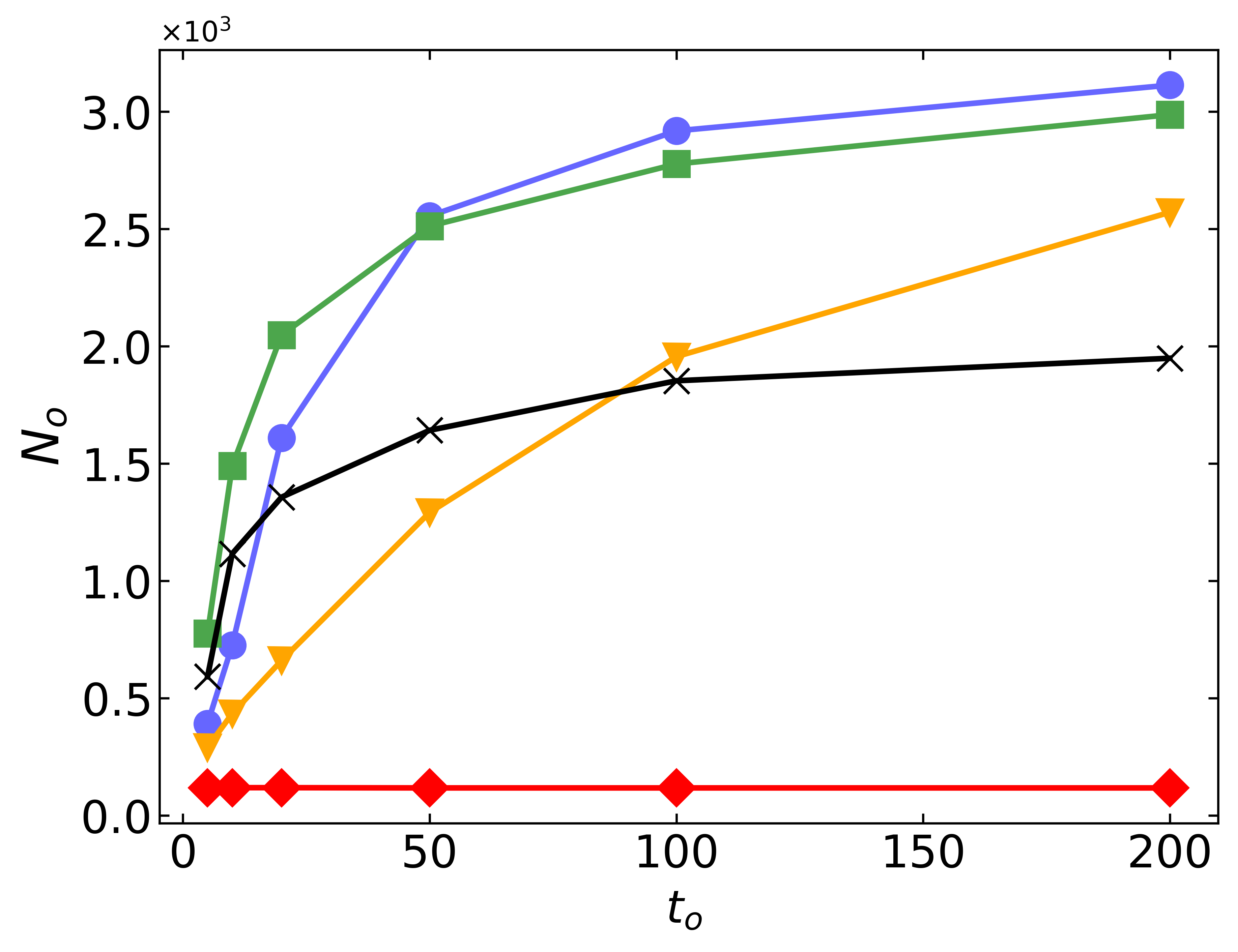}
  \caption{Number of relevant Reflection operators, for  different integration times $t_{0}$ and dynamics. Same color code as in Fig. \ref{fig6}. In this case we have used $N=2^{6}+1$ as the dimension of Hilbert space.}
  \label{fig8}
\end{figure}

On the other hand, it is interesting to investigate if the amount of relevant operators changes as a function of 
the integration time $t_{0}$, and eventually how this change is. In Figures \ref{fig6}, \ref{fig7} and \ref{fig8} we 
show the number of relevant operators for Pauli, translation and reflection bases respectively for each dynamics and different integration times, needed to achieve the effective $S_{L}$ behavior. For all bases, we notice that if the system 
has at least one 
hyperbolic degree of freedom, the number of operators grows steeply with the integration time. If the system is 
completely elliptic and the coherent states are located at the fixed point, the number of operators is essentially 
constant, while if they are not at a periodic orbit, the number of operators grows with a rate much slower than in the 
mixed (HE and EH; we have taken both points of view in order to better look into dynamical properties) 
or totally hyperbolic (HH) cases, specially for the reflection base (see Fig. \ref{fig8}). 
For the HH case, if we take long integration times, we will have that almost all operators (not all since we only require 
the effective $S_{L}$ behavior) are relevant and equivalent 
reminding us of the underlying classical ergodicity in this scenario. 
Growth in the number of relevant operators gives us more hints on the OTOCs sensitivity for quantum complexity, providing with an alternative natural indicator of it. 
As a final remark, from Figs. \ref{fig6}, \ref{fig7} and \ref{fig8} we see that the number of relevant operators can be 
dependent on the base. An extreme example is given by the Kirkwood one whose operators are defined by
\begin{equation}
  K_{(i,j)}=\ket{q_i}\bra{p_j}, 
\label{eq:Kirk}
\end{equation}
and for which there is a clear association with phase space representations, having a direct classical meaning \cite{lakshminarayan2018out}. 
For any of the operators in this basis it is straightforward to show that 
\begin{equation}
  C_{K(i,j)}(t)= \rho_A^2(t),
\label{eq:Kirk2}
\end{equation}
hence all of them are equally relevant in sensing the dynamics, so special care must be taken at the time 
of selecting the base if one wants to profit from the OTOCs ability to characterize quantum complexity.

\begin{figure}[!htp]
\centering
  \includegraphics[scale=0.5]{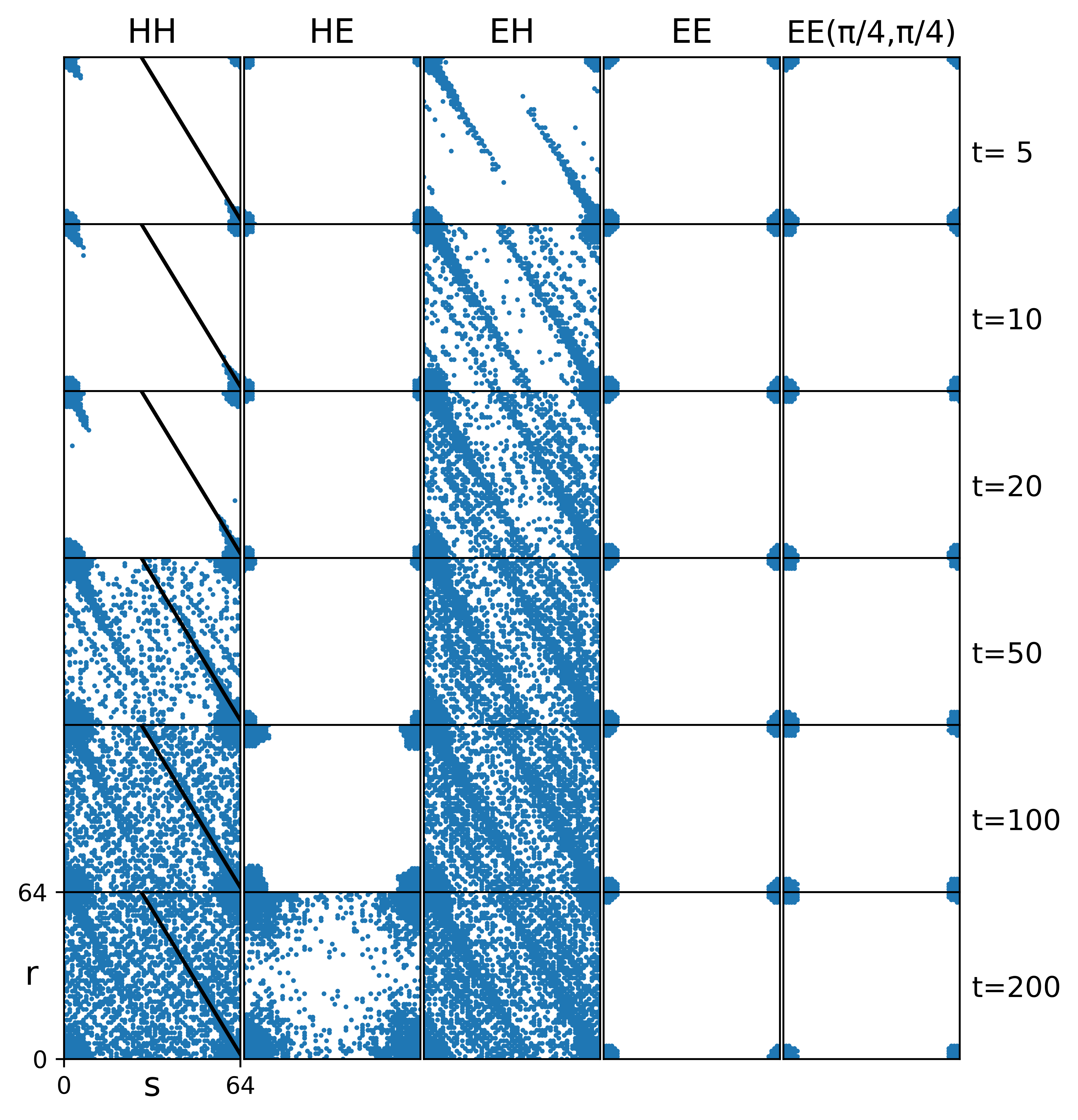}
	\caption{Relevant translation operators in phase space, for each dynamics as we increase the integration time $t_{0}$. Each point $(r,s)$ represents a translation chord, where $r$ indicates the translation in momentum and $s$ in position. The black solid line represents the unstable manifold direction of our map. $N=2^{6}$.}
  \label{fig9}
\end{figure}

\begin{figure}[!htp]
\centering
  \includegraphics[scale=0.5]{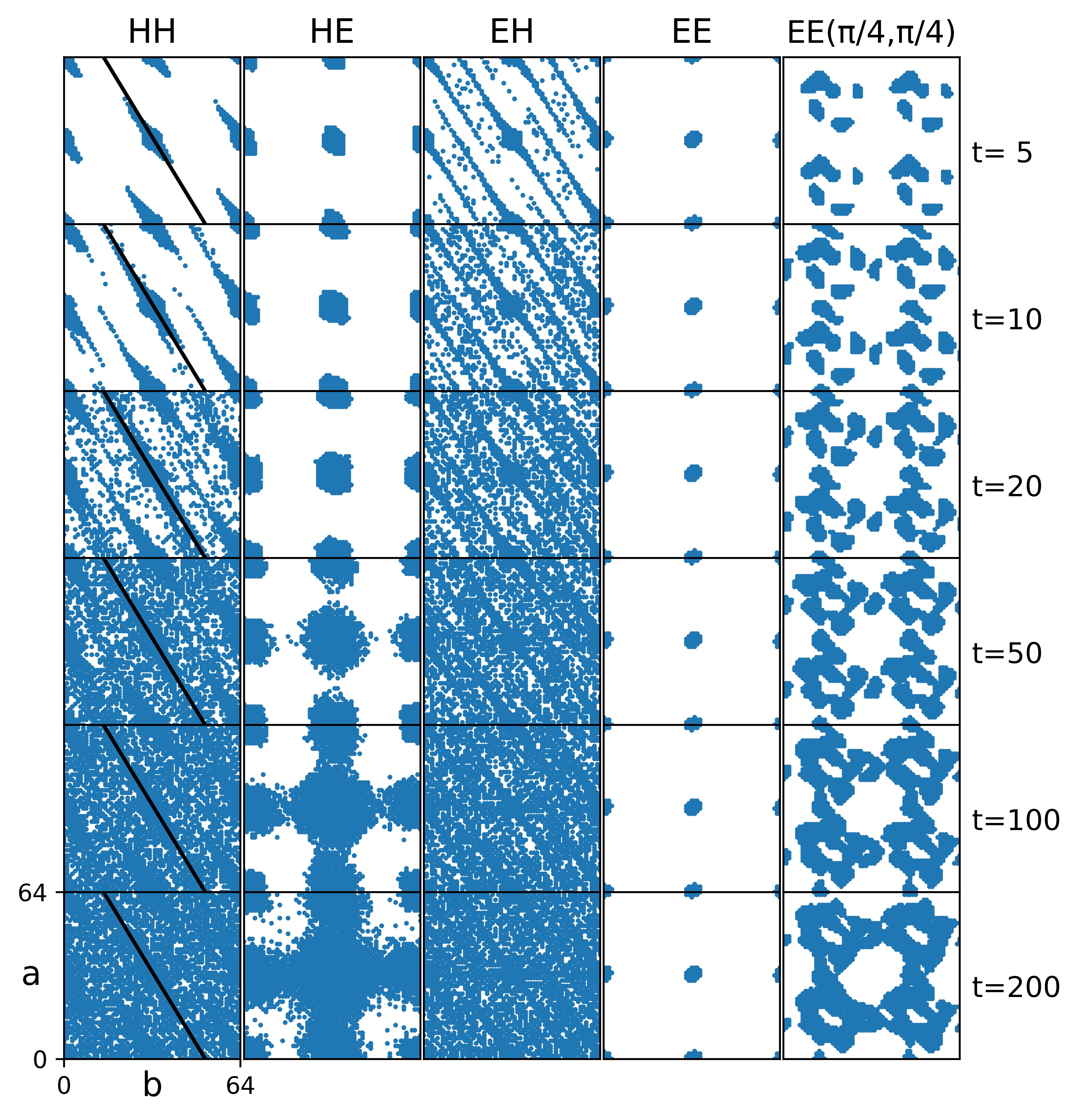}
  \caption{Relevant reflection operators in phase space, for each dynamics as we increase the integration time $t_{0}$. Each point $(a,b)$ represents a reflection center, where $a$ indicates its momentum and $b$ its position. The black solid line represents the unstable manifold direction of our map. 
  In this case we have used $N=2^{6}+1$ as the dimension of Hilbert space.}
  \label{fig10}
\end{figure}

All the previous analysis has led us to look for an explanation on the physical meaning of the operator relevance 
at the time to describe the $S_L$ behavior or the quantum complexity in general. 
In order to proceed we restrict ourselves to translation and reflection operators since they can be represented in chord and center phase space. In Eqs. \ref{eq:TRAbase} and \ref{eq:REFbase}, we identified each one of these operators with a couple of indexes, $(r,s)$ for translations and $(a,b)$ for reflections, which are related to the chord of translation and the reflection center, respectively. These indices can be represented in a $2D$ plot, allowing to visualize the different operators. Figures \ref{fig9} and \ref{fig10} show the most relevant translation and reflection operators for each dynamics and different integration times $t_{0}$. For reflections, we have chosen an odd Hilbert space dimension of $N=65$ in order to deploy \cite{rivas1999weyl} the complete basis from the quarter torus with half integer indices into the full one with integer indices. This allows a clearer visualization  of the classical structures in phase space \cite{rivasscars}.

In the HH case, we see that the relevance of translation and reflection operators grows along the unstable manifold of our map, 
indicated in Figures \ref{fig9} and \ref{fig10} with a black solid line. The number of relevant operators grows with $t_0$ and 
finally extends to the entire phase space.
In the HE scenario, the relevant translation operators (see Fig. \ref{fig9}) are grouped around the identity operator 
(the chord is null) because we are looking at the elliptic degree of freedom. However, the number of them increases 
with $t_0$, reflecting the spreading of the coherent state 
due to the influence of the hyperbolic map. Relevant reflections are concentrated in the center of the 
phase space where the coherent state is (see Fig. \ref{fig10}) and, when $t_0$ increases more operators are 
needed to describe the dynamics enlarging the corresponding distribution. 
A similar situation arises in the EH case (now we 
observe the hyperbolic subsystem), i.e. the number of relevant operators grows and its distribution spreads along the 
unstable direction for both translations and reflections. Finally, in the EE case for the coherent state at the periodic point, the distributions remain localized and again the translation operators which better capture the dynamics are the ones 
closer to the identity, while the corresponding reflection operators are those at the center of the phase space. 
In both cases the number of relevant operators does not change. Finally, if we locate the coherent states out of the fixed point, the relevant translation operators are still closer to the identity but the reflection ones follow the evolution of 
the distributions as in \cite{bergamasco2017}.

In all these cases, we can observe that the set of relevant operators follow the footprints of the classical dynamical 
evolution and this provides with a clear interpretation of the relevance criterion developed in this work. However, 
we underline that some bases of operators are more sensitive than others, following the footprints closer and allowing to 
reveal the classical structures and the quantum complexity in a clearer way.
As we have previously seen in Eq. (\ref{eq:Kirk2}), for the Kirkwood base all of the OTOCs are equivalent in following 
the linear entropy behavior. For a pure state $\hat{\rho} $ with the translation operators basis, we can see that
the OTOC can be expressed in terms of 
\[
\rho_\xi(t) = Tr(\hat{T}_\xi \hat{\rho}(t) ),
\]
 the chord representation of the evolved density $\hat{\rho}(t) $, as
\begin{equation}
 C_{T \xi(t)}= \rho_\xi(t) \rho_{-\xi}(t).
\label{eq:trans}
\end{equation}
Meanwhile, for the reflection basis, the OTOC can be expressed in terms of the Wigner function 
\[
W_x(t) = (2 \pi \hbar) \  Tr(\hat{R}_x \hat{\rho}(t) ),
\]
as
\begin{equation}
  C_{R x(t)}= \frac{1}{(2 \pi \hbar)^2} W_x^2(t).
\label{eq:refle}
\end{equation}
This makes the OTOC in the reflection basis remarkably sensitive to the classical structures in phase space, providing 
with a very clear link to complexity measures \cite{Benenti-Carlo-Prosen}. 

\section{Conclusions}
\label{sec:conclusion}

Recently quantum chaos and high energy physics have become closely related through a chaoticity measure, the OTOC. An interesting 
bridge towards a more general interpretation as a complexity measure has been provided from the quantum information perspective 
via the OTOC-RE theorem which relates it to the second Renyi entropy \cite{fan2017out,hosur2016chaos,PhysRevResearch.1.033044}. 
In this work we have deepen on the study of this relation for a paradigmatic bipartite system covering the main kinds of dynamics, 
i.e. two coupled and perturbed Arnold cat maps. We have studied the behavior of three different bases of operators, 
namely the Pauli, translation and reflection ones.

We have defined a criterion of relevance for each operator from these bases relying on their corresponding OTOC contribution to 
the linear entropy $S_L$ up to time $t_{0}$. Armed with this tool we have found that less than $35\%$ of the operators of these  widely used bases are enough to reach the effective $S_L$ behavior. This means that to characterize the system in terms of its complexity the whole basis of operators 
is not needed in general but a much lower fraction instead (we underline that this is basis dependent though). 
The least relevant operators 
revealed as poor indicators of the dynamical complexity of the system. Moreover, the scaling of the number of relevant operators as a 
function of the time $t_0$ proved to be an alternative indicator of complexity, much in the same sense as the scaling of the 
number of operations is a measure for algorithmic complexity.

Finally, for the translation and reflection operators which can be directly represented in phase space our concept of 
relevance turns out to have an easy interpretation. The set of relevant operators follows the quantum footprints of 
the corresponding classical evolution (more or less closely depending on the basis).
In the future we will investigate this relation even more deeply taking into account generic density operators.

\end{document}